\documentclass[useAMS,usegraphicx,usenatbib]{mn2e}
\usepackage{txfonts,amssymb,hyperref,subfigure,epsfig}

\defcitealias{2004ApJ...604..388I}{IL1}
\defcitealias{2004ApJ...616..567I}{IL2}
\defcitealias{2005ApJ...626.1045I}{IL3}

\title[Planet Formation around Brown Dwarfs]{The potential for Earth-mass
  planet formation around brown dwarfs} 
\author[M.~J.~Payne and G.~Lodato]{Matthew~J.~Payne$^{1}$
  \thanks{E-mail:~\href{mailto:mpayne@ast.cam.ac.uk}{mpayne@ast.cam.ac.uk}
    (MJP)} and Giuseppe~Lodato$^{2,1}$\\
  $^1$Institute of Astronomy, University of Cambridge, Madingley Road, 
  Cambridge CB3 0HA\\
  $^{2}$Department of Physics and Astronomy, University of Leicester,
  Leicester, LE1 7RH}

\begin{document}

\date{Accepted ---. Received ---; in original form ---}

\pubyear{2007}

\maketitle

\label{firstpage}

\begin{abstract}

  Recent observations point to the presence of structured dust grains in the
  discs surrounding young brown dwarfs, thus implying that the first stages of
  planet formation take place also in the sub-stellar regime. Here, we
  investigate the potential for planet formation around brown dwarfs and very
  low mass stars according to the sequential core accretion model of planet
  formation. We find that, for a brown dwarfs mass $0.05M_{\odot}$, our models
  predict a maximum planetary mass of $\sim 5 M_{\oplus}$, orbiting with
  semi-major axis $\sim 1 \textrm{AU}$. However, we note that the predictions
  for the mass - semi-major axis distribution are strongly dependent upon the
  models chosen for the disc surface density profiles and the assumed
  distribution of disc masses. In particular, if brown dwarf disc masses are
  of the order of a few Jupiter masses, Earth-mass planets might be relatively
  frequent, while if typical disc masses are only a fraction of Jupiter mass,
  we predict that planet formation would be extremely rare in the substellar
  regime. As the observational constraints on disc profiles, mass dependencies
  and their distributions are poor in the brown dwarf regime, we advise
  caution in validating theoretical models only on stars similar to the Sun
  and emphasise the need for observational data on planetary systems around a
  wide range of stellar masses.  We also find that, unlike the situation
  around solar-like stars, Type-II migration is totally absent from the planet
  formation process around brown dwarfs, suggesting that any future
  observations of planets around brown dwarfs would provide a direct measure
  of the role of other types of migration.
\end{abstract}


\section{Introduction}\label{Introduction}
Observations of planets proceed apace: currently there are 232 extra solar
planets known in 199 different planetary systems\footnote{http://exoplanet.eu,
  06/05/07}. As is well known, many, if not all, of these systems are
radically different to the Solar System \citep{2004MNRAS.354..763B}, often
posing challenges to theories of planet formation. As time has progressed, a
number of planets have begun to be observed around low mass stars
($M_{\star}\sim 0.2 - 0.3 M_{\odot}$), thus begging the natural question of
whether this trend will continue and whether similar planetary systems are to
be expected at even lower masses, down in the brown dwarf regime.

Observations suggest that there does not seem to be any discontinuity between
the properties of the discs around stars and those around brown dwarfs:
Spitzer observations by \citet{2005ApJ...631L..69L} found that disc fractions
for stars and brown dwarfs in the IC 348 and Chamaeleon I clusters were
similar, thus suggesting a similar disc lifetime (see also
\citealt{2004A&A...427..245S}).  The measurement by
\citet{2006ApJ...645.1498S} of the Taurus star-forming region indicates that
many of the discs around brown dwarfs must be relatively large in size,
extending beyond $10\textrm{AU}$, and have masses up to 12\% of the central
brown dwarf, with the average being 2.6\%, compared with an average value of
1.9\% for low mass main-sequence stars (note, however, that such disc mass
estimates are very uncertain). Hence it appears that discs around brown dwarfs
have similar properties and are as ubiquitous as discs around main sequence
stars.

Observations indicate that grain growth, the precursor to planet formation in
the sequential core accretion model, can occur efficiently in the small discs
surrounding brown dwarfs \citep{2005Sci...310..834A}. This then poses such
questions as: what is the potential to form planets around very low mass
stars; what would be their mass-semi-major axis distribution and would they be
detectable? If formed or forming planets could be observed around brown
dwarfs, these would provide highly valuable, low mass calibration points for
the assessment of planet formation models over a wide range of stellar masses.
In view of this, and in view of forthcoming missions such as the Terrestrial
Planet Finder and Darwin, aimed at looking for Earth like planets, it is
important to estimate the likelihood of the presence of planets around stars
of different masses, and in particular around brown dwarfs.

Planet formation models currently fall into two categories: (i) the sequential
core accretion model, which can account for the formation of both rocky
planets and gas giants and (ii) the gravitational instability model, which can
only account for the formation of massive gas giants. The sequential core
accretion model of planet formation \citep{1996Icar..124...62P} assumes that
planets form through a process in which first dust grains initially present in
the protoplanetary disc condense to form kilometre-scale planetesimals; the
collisional accumulation of planetesimals forms the basis for the runaway
growth of a protoplanetary core of rock and/or ice
\citep{1972book.S,1980ARA&A..18...77W}, and finally if a critical mass has
been reached, the protoplanetary cores can begin to rapidly accrete a gaseous
envelope \citep{1980PThPh..64..544M, 1986Icar...67..391B,
  1996Icar..124...62P}. The gravitational instability model does not need to
form the initial rocky core, and assumes that the protoplanetary disc may
become gravitationally unstable and hence form gas-giants via gravitational
collapse of high density regions in the disc
\citep{1998Natur.393..141B,2000ApJ...536L.101B,2004ApJ...609.1045M}. Whilst it
is possible that some of the observed planets formed via a
gravitational-instability scenario (and indeed also planetary mass companions
to brown dwarfs. See, for example, the case of 2MASS1207B,
\citealt{2005A&A...438L..25C,2005MNRAS.364L..91L}), the results of
\citet{2007astro.ph..3237M} suggest that around 90\% of the observed
extra-solar planets have a mass - semi-major axis relation consistent with the
core-accretion model and that the metallicity dependence of the distribution
of these planets can only be explained using the core-accretion model.

\begin{figure}
  \includegraphics[origin=c,width=\columnwidth,trim=0 0 3
  3,angle=0,clip=true]{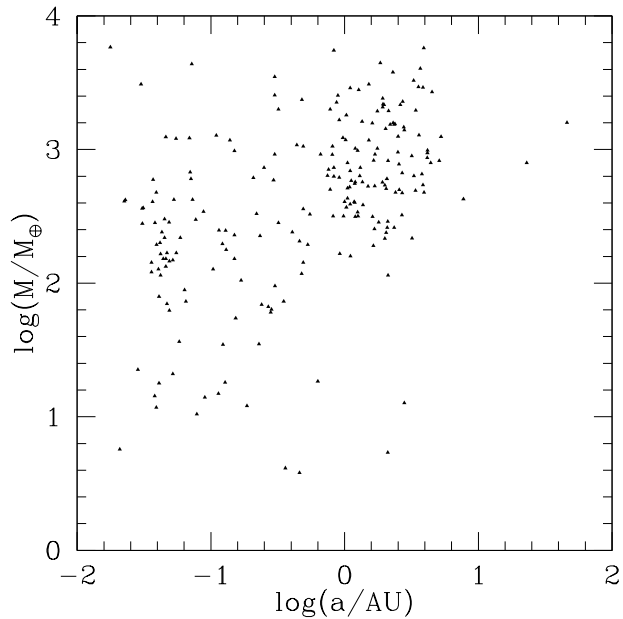}
  \caption{Mass - Semi-major axis distribution of all currently known
    extra-solar planets.}
\label{FIG:EXOPLANETDATASIMPLE}
\end{figure}

In a series of papers, \citet[][henceforth \citetalias{2004ApJ...604..388I,
  2004ApJ...616..567I, 2005ApJ...626.1045I}]{2004ApJ...604..388I,
  2004ApJ...616..567I, 2005ApJ...626.1045I} attempted to construct a
semi-analytic model of the combined core accretion and migration process which
would allow quantitative comparison to be made between the observed
distribution of extra-solar planets and that predicted by the theoretical core
accretion model. They concentrated on investigating the observed mass -
semi-major axis distribution of extrasolar planets, shown here in Fig.
\ref{FIG:EXOPLANETDATASIMPLE}. This model starts with kilometre-size
planetesimals, embedded in a two component gas and dust disc. The subsequent
core accretion of planetesimals onto the protoplanet is modeled using the
combined results of analytic models and N-body simulations. Once the cores
reach a given critical mass, gas accretion is assumed to start, this process
being modeled using numerical results for Kelvin-Helmholtz contraction of the
gas envelope. This model also uses a number of prescriptions to enforce the
limitation to core growth caused by various effects, such as isolation and
migration.

Whilst this kind of modeling is admittedly oversimplified, the models of
IL1-IL3 do succeed in reproducing several features of the observed
distribution of extra-solar planets, such as the metallicity dependence for
the probability of planet formation, the dependancy on stellar mass of the
distribution of planetary masses as well as giving realistic mass ranges for
the outcome of the planet formation process. However, the models experience
problems in reproducing observations of intermediate mass planets with
semi-major axes less than 3 AU. It is possible that these difficulties stem
from the neglect of Type I migration in these models - see sections
\ref{Planetary Migration} and \ref{Planet Formation around Intermediate-Mass
  Stars} for further consideration.

Clearly a self-consistent disc processing and core growth model for multiple
interacting planets would be more appropriate (e.g.
\citealt{2004ApJ...612L..73L, 2005A&A...433..247P, 2005A&A...434..343A,
  2006RPPh...69..119P, 2007ApJ...654.1110T}). But given that the models of
IL1-IL3 are currently the only extant attempt at providing a general framework
for examining the outcome of the planetary formation and migration mechanism,
they provide a useful framework for the exploration of the extensive parameter
space and they allow us to start examining in a statistical manner the
potential for planet formation in discs around brown dwarfs and the potential
for practical observations in the near future.

To address the issue of planet formation around brown dwarfs under the core
accretion model, we therefore extend the core-accretion model of IL1-IL3 down
to the low mass regime. Once the basic structure of the model is in place, it
can be used to investigate the effect on the planetary mass - semi-major axis
distribution of varying key physical parameters. These include: the surface
density profile of the disc, the size and mass of the disc, the rate of disc
clearance, the effect of stellar mass and luminosity as well as migration.

\section{Planet Formation Model}\label{Planet Formation Model}
The model we adopt here is primarily based on the model of sequential
accretion presented in \citetalias{2004ApJ...604..388I} -
\citetalias{2005ApJ...626.1045I}. In this section we initially briefly recount
the principle components which pertain to our investigation before going on to
discuss in more detail the refinements and additions which are required to
extend the model down to the brown dwarf regime. Readers requiring more detail
on the original model construction should refer to
\citetalias{2004ApJ...604..388I} - \citetalias{2005ApJ...626.1045I}.

\subsection{Disc Properties and Initial Conditions}\label{Disc Properties and
  Initial Conditions}
The protoplanetary disc is modeled using a two component surface-density
profile, $\Sigma = \Sigma_d + \Sigma_g$, to account for the dust and gas,
with:
\begin{eqnarray}
  \Sigma_d &=& 10 f_{0} \eta_{ice}  \left(\frac{a}{1 AU}\right)^k 
  \left(\frac{M_{\star}}{M_{\odot}}\right)^n \textrm{g cm}^{-2} \\
  \Sigma_g &=& 2.4 f_{0} \times 10^{3}  \left(\frac{a}{1 AU}\right)^k 
\left(\frac{M_{\star}}{M_{\odot}}\right)^n e^{-t/\tau_{disc}} \textrm{g cm}^{-2}.
\label{eq:1}
\end{eqnarray}
where $M_{\star}$ is the mass of the central star, $a$ is the radial distance
from it, and $\eta_{ice}$ is a step-function accounting for the increase in
dust density at the snow line such that $\eta_{ice}(a<a_{ice})=1$ and
$\eta_{ice}(a>a_{ice})=4.2$. The radial location of the snow-line $a_{ice}$ is
set by the stellar luminosity, $L_{\star}$, such that
\citep{1981PThPS..70...35H}:
\begin{eqnarray}
a_{ice} = 2.7 \left(\frac{L_{\star}}{L_{\odot}}\right)^{1/2} \textrm{AU.} 
\end{eqnarray}

The factors $\left(M_{\star}/M_{\odot}\right)^n$ allow an investigation of the
dependence of the disc mass on stellar mass (here we examine the cases $n=1
\textrm{ and }2$). The dissipation of the gaseous component is modeled as a
simple exponential decay with characteristic timescale, $\tau_{disc}$, chosen
to be $\sim 10^6 - 10^7$ yrs, as suggested by observations of the infrared
excess in young stellar objects \citep{haisch01}. The dust component is
assumed to be time-independent. Setting the parameters $f_0 = 1$ and $k=-3/2$
would give the minimum mass nebula model (MMNM) \citep{1981PThPS..70...35H}.
We investigate the effect of varying the absolute surface density of the disc
by varying $f_{0}$.

While \citetalias{2004ApJ...604..388I} - \citetalias{2005ApJ...626.1045I}
only consider the case where the surface density profile is a power law with
index $k=-3/2$, here we also investigate the effects of having $k=-1$ (see
section \ref{Surface Density Dependence on Semi-Major Axis} for motivation and
discussion).

It should also be noted that the simple exponential decay of the gaseous disc
is the only evolution of the disc to be considered. The solid component
is taken to be constant and neither component is modeled self-consistently
with the growing mass of the planet. A self-consistent disc processing and
core growth model would clearly be more appropriate (e.g.
\citealt{2005A&A...434..343A}), but would also be more computationally
demanding, so we adopt the simplified approach in this investigation.

To investigate the formation of planets around brown dwarfs, it is clear that
we need to extend the semi-analytic model down to $M_{\star}<0.084M_{\odot}$.
However, the model depends on $M_{\star}$ in two main ways: (i) Direct
dependence on $M_{\star}$ via $\Sigma\propto M_{\star}^n$; and (ii) Indirect
dependence on $M_{\star}$, e.g.  the dependence of $a_{ice}$ and hence
$\Sigma_d$ on $L_{\star}$. The detailed effects of varying $L_{\star}$ are
discussed below in sections \ref{Luminosity Evolution} and \ref{sec:lum}.

\subsection{Core Growth and Gas Accretion}\label{Core Growth and Gas
  Accretion}

We follow the prescription for core growth and gas accretion as presented in
\citetalias{2004ApJ...604..388I}, with a core accretion timescale for a planet
of mass $M$ of \citep{2002ApJ...581..666K}
\begin{eqnarray}
  \tau_c &=& 1.2 \times 10^5 \left(\frac{{\Sigma}_d}{10 \mbox{g
        cm}^{-2}}\right)^{-1} 
\left(\frac{a}{1 AU}\right)^{3/5} \left(\frac{M}{M_{\oplus}}\right)^{1/3} 
  \left(\frac{M_{\star}}{M_{\odot}}\right)^{-1/6} \nonumber
  \\ 
  && \times \left(\frac{{\Sigma}_g}{2.4 \times 10^3 \mbox{g
        cm}^{-2}}\right)^{-2/5} 
\left(\frac{m}{10^{18} g}\right)^{2/15} yr. \label{taucfullDetailed}
\end{eqnarray}
where $m=10^{18}\textrm{g}$ is the assumed mass of the planetesimals accreted
by the planetary core.

\citet{2000ApJ...537.1013I} find that once a critical core mass of
\begin{eqnarray}
M_{\rm crit}= 10 \left(\frac{\dot{M}}{10^{-6}M_{\oplus}\textrm{yr}^{-1}}\right)^{0.25} M_{\oplus}
\label{Mcrit} 
\end{eqnarray}
has been reached, gas accretion can commence at a rate of 
\begin{eqnarray}
\tau_{K-H} &=& 10^{10} \left(\frac{M}{M_{\oplus}}\right)^{-3} yr. \label{tauKH}
\end{eqnarray}

Once a planet has grown to such as size that it opens a gap in the disc, the
gas accretion rate is regulated by the rate at which the disc can process
material and supply it across the gap to the planet.
\citetalias{2004ApJ...604..388I} use a mass transfer rate given by the
inferred rate for protostellar discs on T Tauri stars
\citep{1998ApJ...495..385H, 2000prpl.conf..377C}
\begin{eqnarray}
  \dot{M}_{disc} &=& 10^{-4} M_{\oplus}
  \left(\frac{t}{\tau_{disc}}\right)^{-3/2} \textrm{yr}^{-1}. 
\label{mdotgap}
\end{eqnarray}
However, once the gap is opened, the accretion rate is reduced to values of
the order of $=0.1\times\dot{M}_{disc}$ \citep{lubow96,lubow06}. Here we have
considered both the case described in Eq. (\ref{mdotgap}), appropriately
scaled down for brown dwarfs, and the case where accretion is further reduced
due to gap opening, and found that this had little effect on the final
distribution of planet masses and semi-major axes.  Indeed, the details of gas
accretion do not strongly affect our conclusions, since in most cases, as
discussed below, we observe only very limited gas accretion in brown dwarf
protoplanets.

We again acknowledge that these growth models are probably oversimplified. The
form for $\tau_c$ given in Eq. (\ref{taucfullDetailed}) neglects the effects
of core migration on the accretion rate of planetesimals onto the
protoplanetary core (e.g. \citealt{2005AJ....130.2884M,ricearmi03}). The
treatment of gas accretion giving rise to the expressions for $M_{\rm crit}$
and $\tau_{K-H}$ in Eqs. (\ref{Mcrit} and \ref{tauKH}) is also highly
approximate, neglecting the effects of opacity \citep{2005AN....326..913A,
  2000ApJ...537.1013I} and all the details evidenced by a numerical treatment
of the problem \citep{2000ApJ...544..481B, 1996Icar..124...62P}.

\subsection{Luminosity Evolution}\label{Luminosity Evolution}
The models of \citetalias{2004ApJ...604..388I} -
\citetalias{2005ApJ...626.1045I} treat the central star as a purely static
object, with a fixed luminosity $L_{\star}$ equal to that on main-sequence,
scaling such that $L_{\star}/L_{\odot}=(M_{\star}/M_{\odot})^{1/4}$. For brown
dwarfs, a constant luminosity model would be completely inappropriate, as no
constant luminosity is ever reached, and we need to take this into account.

\begin{figure}
  \includegraphics[origin=c,width=\columnwidth,trim=0 0 3
  3,angle=0,clip=true]{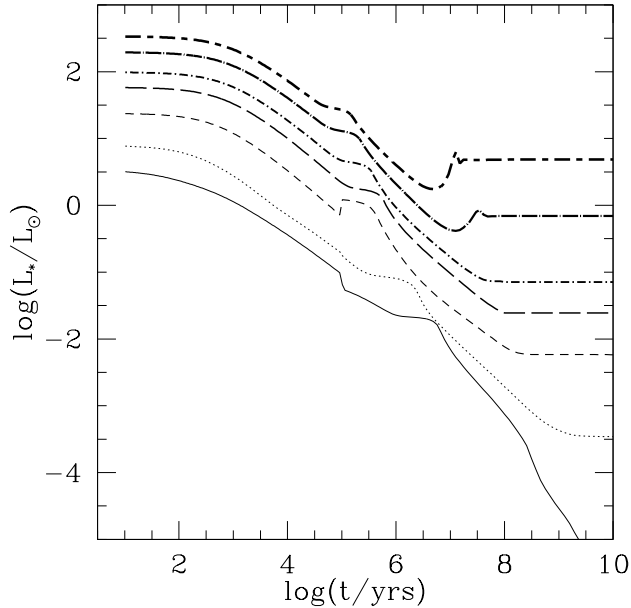}
  \caption{Luminosity evolution profiles from \citet{1999MNRAS.310..360T} for
    (top-to-bottom) $M_{\star}/M_{\odot}=$ 1.5, 1.0, 0.6, 0.4, 0.2, 0.084 
    and 0.05
    respectively.}
\label{FIG:LUM:EVO}
\end{figure}

We allow stellar luminosity evolution to take place, accounting for this using
the pre-main-sequence evolution simulations of \citet{1999MNRAS.310..360T}.
Some sample evolutionary tracks are shown in Fig. \ref{FIG:LUM:EVO}. Compared
to a static luminosity model, the net effect of the inclusion of luminosity
evolution is to \emph{raise} the luminosity at earlier times.

It should be noted that the core accretion model considered here, with an
initial swarm of planetesimals of mass $10^{18} \textrm{g}$, may have required
$\sim10^4\textrm{yrs}$ for the planetesimals to form, although theoretical
models vary considerably
\citep{1993prpl.conf.1031W,1997Icar..129..539S,riceetal04,2006MNRAS.372L...9R}.
Hence we set the zero of time in our model to correspond to $10^4
\textrm{yrs}$ in Fig.\ref{FIG:LUM:EVO}.

\subsection{Planetary Migration}\label{Planetary Migration}

Protoplanetary cores embedded within a gaseous disc interact with the gas and
are thus subject to gravitational torques which cause them to migrate away
from their initial position. For small core masses the interaction is linear
and the corresponding migration regime is called Type I, while for larger core
masses (typically above 1 Jupiter mass, for planets forming around a solar
type star), a gap is opened in the disc and the resulting migration regime is
called Type II \citep{1997Icar..126..261W}. 

Type I migration is important for a number of reasons: (i) It is well known
\citep{2006RPPh...69..119P} that the inclusion of Type-I migration in models
of planet formation around solar-mass stars completely prevents the formation
of massive gas giants (unless the disc mass is sufficiently small
\citealt{thommes06}), as all the planetary cores rapidly migrate into the
central star, and thus Type I migration is generally neglected. However,
simulations by \citet{2004MNRAS.350..849N} suggest that, in a turbulent disc,
Type-I migration may actually follow a random walk, potentially providing a
mechanism to ameliorate this problem. (ii) It may help to explain the presence
of intermediate mass planets at small semi-major axes, by providing an
additional migratory mechanism. In this paper, we have run models both
including and neglecting the effect of standard \citep{1997Icar..126..261W}
Type-I migration.

Type-II migration occurs when the planet becomes massive enough to open a gap
in the disc, the subsequent orbital evolution of the planet becoming firmly
coupled to the viscous evolution of the disc \citep{1985prpl.conf..981L}. A
gap is opened up when the planetary mass exceeds
\begin{eqnarray}
  \frac{40\nu}{a^2\Omega_K}M_{\star} &=& 40\alpha\left(\frac{c_s}{V_K}\right)^2 M_{\star} \nonumber
\\
 &=& 3 \left(\frac{\alpha}{10^{-4}}\right) \left(\frac{a}{1 AU}\right)^{1/2} \left(\frac{L_{\star}}{L_{\odot}}\right)^{1/4} M_{\oplus}, \label{Mgvisfinal}
\end{eqnarray}
where we have used the $\alpha$-parameterization for the effective viscosity,
$\nu=\alpha H^2 \Omega_K$, of \citet{1973A&A....24..337S} and we use $c_s^2
\propto T$.

However, hydrodynamical simulations
\citep{1985prpl.conf..981L,2004ApJ...604..388I} suggest that the onset of
Type-II migration occurs somewhat later, when the planet has grown to a larger
mass. Type-II migration is therefore taken to begin when the planet exceeds 10
times the gap opening mass.

In general, in any viscously evolving gaseous disc, the inner parts move in
and accrete, while the outer parts move out and take up the angular momentum
lost by the accretion material \citep{1974MNRAS.168..603L}. The direction of
Type II migration is therefore determined by the position of the planetary
core with respect to the radius which marks the transition from the inner to
the outer disc, which we obtain from the appropriate self-similar solution of 
\citet{1974MNRAS.168..603L}.

\subsection{Methodology}\label{Methodology}
Given the above semi-analytic prescription for protoplanetary growth, if we
specify $M_{\star}$, $k$, $n$, $f_{0}$, $\tau_{disc}$ and the initial
semi-major axis $a_i$, we can then allow the planet to grow from an initial
mass of $10^{22}$ g. If we then perform a Monte-Carlo simulation, allowing the
parameters $f_{0}$, $\tau_{disc}$ and $a_i$ to vary, we can build up a picture
of the potential for planet formation for a wide range of protoplanetary discs
around a wide range of stars.

Because the initial size distribution is unknown for brown dwarf disks, the
dissipation timescale is also unknown. Attempts at estimating this timescale
have been made by \citet{2006ApJ...639L..83A} based on the $\dot{M}-M$
relation \citep{2004A&A...424..603N}, and they argue that $t \sim
M_{\star}^{-1}$. However, it is not clear whether such observational scalings
really reflect some intrinsic properties of the initial protostellar disc
population, nor how much are they affected by selection effects
\citep{2006MNRAS.370L..10C}, so we have kept the same timescale for solar mass
stars and brown dwarfs (but we do allow for a relatively large spread in
timescale, as suggested by \citet{2007MNRAS.376.1350C}). We therefore allow
$\tau_{disc}$ to be evenly distributed in logarithmic scale over $10^{6} -
10^{7} \textrm{yrs}$.

We take $f_0$ to be log-normally distributed (centred on $\log f_0 =0$ and
with variance equal to 1) and then demand that the disc mass within $100
\textrm{AU}$ be less than $0.35M_{\star}$, to ensure that the disc is
gravitationally stable out to $100\textrm{AU}$. We thus end up with a
distribution of disc masses such that the maximum disc/stellar mass fraction
is $\leq 0.35$, with an average significantly below this value. Note that
\citet{2005ApJ...621L.133R} have observed protostellar disc masses as high as
$0.35M_{\star}$, indicating that the above upper limit is reasonable. Finally,
we distribute the initial orbital semi-major axes evenly in log scale over
$0.1$ to $100$ AU and terminate all migration at $0.04$AU\footnote{While in
  general, for solar-type stars, this radius corresponds to the stellar
  magnetospheric radius, there is no strong reason for assuming that the
  termination radius in brown-dwarf discs would be the same, but as we show
  later, we find little or no migration around brown dwarfs, so its exact
  value is effectively unimportant.}

\section{Results}\label{Results}

Before discussing our results on planet formation around brown dwarfs, we
initially test our model by reproducing some of the main results of
\citetalias{2004ApJ...604..388I} - \citetalias{2005ApJ...626.1045I} for
planets around Solar-type stars. Our standard model assumes a disc profile
with $\Sigma \propto a^{-3/2}$ and $f_0 \propto
\left(M_{\star}/M_{\odot}\right)^2$ and considers only Type-II migration.  We
then extend our standard model down to the Brown Dwarf regime, adding in also
the effect of luminosity evolution (section \ref{BD}). Finally, we investigate
the impact of varying our main parameters on the potential for planet
formation (section \ref{parameter}).

\subsection{Planet Formation around Solar-Mass Stars}
\label{Planet Formation around Solar-Mass Stars}

\begin{figure}
  \includegraphics[origin=c,width=\columnwidth,trim=0 0 0
  0,angle=0,clip=true]{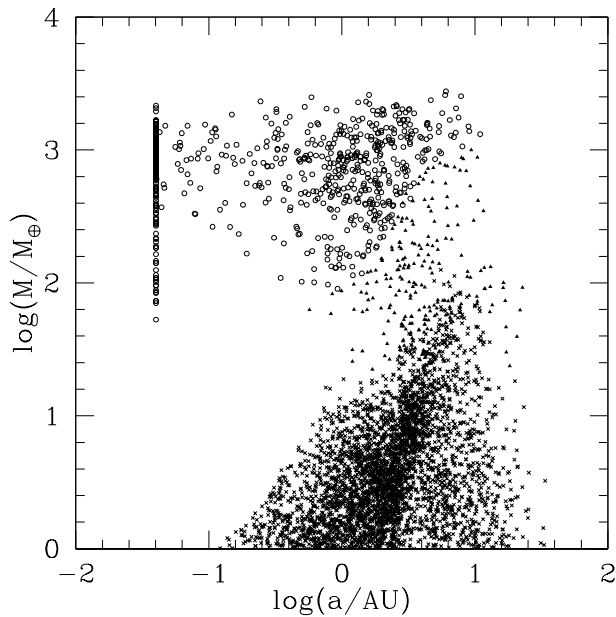}
  \caption{Planet masses $> 1M_{\oplus}$ for $M_{\star}=M_{\odot}$ with static
    luminosity. The different symbols respectively label: Open Circles -
    ``Gaseous'' ($M_g > 10M_c$); Filled Triangles - ``Intermediate
    Composition'' ($10M_c > M_g > M_c$); Crosses - ``Rocky'' ($M_g < M_c$). }
\label{FIG:ILcomparisonEarth}
\end{figure}

Fixing $M_{\star}=M_{\odot}$, we reproduce in Fig.
\ref{FIG:ILcomparisonEarth} the Mass - Semi-Major axis distribution as found
in \citetalias{2004ApJ...604..388I}, fig. 12c. The model uses our standard
values such that $k=-3/2$ in $\Sigma \propto \left(a/1 \textrm{AU}\right)^k$
and the distribution of values of $f_0$ is such that the maximum disc mass is
$0.35M_{\star}$, with an average of $0.05M_{\star}$. We use the following
nomenclature, approximately equivalent to that of
\citetalias{2004ApJ...604..388I}, to label the planets:
\begin{itemize}
\item Gaseous: $M_g > 10M_c$; 
\item Intermediate: $ M_c < M_g < 10 M_c$;
\item Rocky: $ M_g < M_c$;
\end {itemize}
where $M_g$ is the final gas mass and $M_c$ is the core mass, such that the
total planet mass is $M=M_g+M_c$.

The model distribution presented in Fig. \ref{FIG:ILcomparisonEarth}
faithfully reproduces that of \citetalias{2004ApJ...604..388I}, fig. 12c: They
both display the characteristic ``planet desert'' for $a<4\textrm{AU}$ and
$10M_{\oplus}<M<100M_{\oplus}$ and both have a maximum gas-giant mass of $\sim
3000M_{\oplus}$. In addition, the distribution of planet types within the
diagram is very similar (although the precise definition of planet-type-labels
may differ slightly). 

Note the high density of massive planets at $a=0.04\textrm{AU}$ caused by the
inward Type-II migration being arbitrarily halted at this distance. 

Note also that below the $1M_{\oplus}$ line, there is a vast tail of
small-mass planets produced by the simulations; planets which grew only very
slowly and/or were growing in a disc whose mass was very small.

\subsection{Planet Formation around Intermediate-Mass Stars}\label{Planet
  Formation around Intermediate-Mass Stars}

The results of Fig \ref{FIG:ILcomparisonEarth} are in good general agreement
with observed extra-solar planets, the exception being the problematic
observations of intermediate mass planets at small semi-major axes.
\citetalias{2005ApJ...626.1045I} note that their results indicate that
close-in, intermediate mass planets cannot form around F,G and K dwarfs, but
may form around M dwarfs and note that the formation of dynamically isolated
low mass planets around such stars would likely be attributable to Type I
migration or the effect of sweeping secular or mean-motion resonances.

\begin{figure}
  \includegraphics[origin=c,width=\columnwidth,trim=0 0 0
  0,angle=0,clip=true]{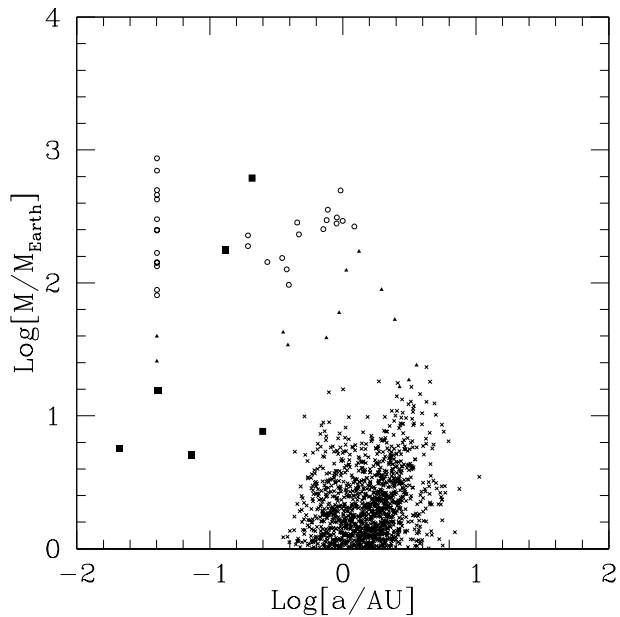}
  \caption{Planet masses $> 1M_{\oplus}$ for $M_{\star}=0.3 M_{\odot}$ with
    only Type II migration present. Labels are as in Fig.
    \ref{FIG:ILcomparisonEarth}, with the addition of observed planets plotted
    as filled squares.}
\label{FIG:03Planets}
\end{figure}

In Fig \ref{FIG:03Planets} we focus on the observed planetary systems around
Gl-581 and Gliese 876, both of which have $M_{\star}\sim 0.3M_{\odot}$. We
thus plot the observed planets on the same scale as our numerical results,
using models that do not include Type I migration. We consistently find it
very difficult to produce close in, intermediate mass planets (such as
Gl-581-c \citealt{2007arXiv0704.3841U}), if we do not include some form of
Type I migration in the process.

However, as mentioned above, the simple introduction of the fast Type I
migration (Ward 1997) would generally preclude the formation of the high mass
($\sim 100 M_{\oplus}$) planets observed in these systems, unless the
migration rate is significantly slower than the standard analytic form
\citep{2004MNRAS.350..849N,thommes06}. It thus appear that, while some
additional form of planetary core migration is needed in order to reproduce
the above-mentioned observations, its rate should be substantially reduced
with respect to the simple analytical estimates. In the rest of the paper we
will therefore consider the two extreme cases in which we neglect completely
Type I migration and in which we include standard Type I migration.

\subsection{Planet Formation around Brown Dwarfs}\label{BD}

If we now allow the central stellar mass $M_{\star}$ to vary down below
$0.084M_{\odot}$, we can to investigate the potential for planet formation in
the brown dwarf regime. As noted in section \ref{Luminosity Evolution}, this
extension requires the implementation of a luminosity evolution model for
brown dwarfs.

\begin{figure}
  \includegraphics[origin=c,width=\columnwidth,trim=0 0 3
  3,angle=0,clip=true]{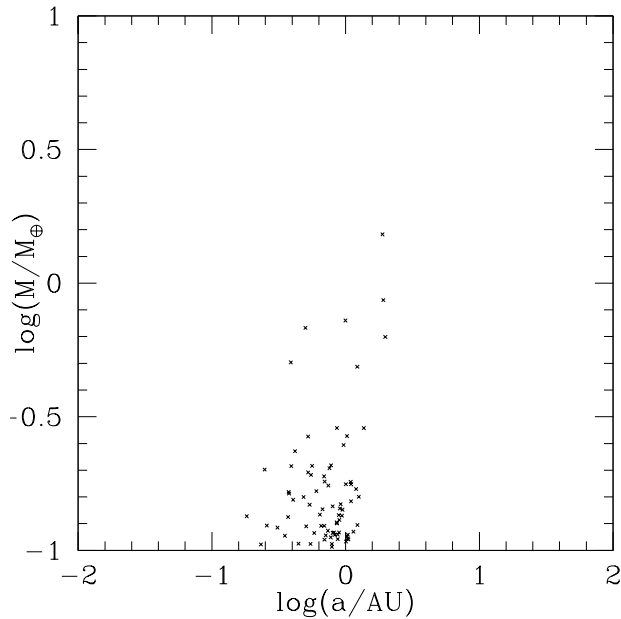}
  \caption{Planet Masses for Brown Dwarf with $M_{\star}=0.05
    M_{\odot}$. Labels are as in Fig. \ref{FIG:ILcomparisonEarth}.}
  \label{FIG:BD1}
\end{figure}

In Fig. \ref{FIG:BD1} we plot the mass - semi-major axis distribution resulting
from our numerical simulation of the planet formation process around brown
dwarfs of mass $0.05M_{\odot}$ - note the change in scale compared to Fig.
\ref{FIG:ILcomparisonEarth}. We keep the standard disc profile of section
\ref{Planet Formation around Solar-Mass Stars} and assume that the disc
surface density scales as $M_{\star}^2$ (i.e., we assume $n=2$ in Equation
(\ref{eq:1})). This effectively makes the $0.35M_{\star}$ cut-off unnecessary,
as the maximum disc mass is now $0.21M_{\star}$ (which corresponds to
$10.5M_{\rm Jupiter}$), while the average disc mass is $0.003M_{\star}$
($0.15M_{\rm Jupiter}$).

Clearly the outcome of the planet formation process around brown dwarfs is
radically different to that for G-dwarfs. Of particular importance is the fact
that all of the planets produced are of low mass - we do not see any
significant number of planets formed above $1M_{\oplus}$, the most massive
ones corresponding to the high tail of the disc mass distribution - and are of
an icy constitution. In none of the simulation runs did any planet grow
massive enough for any significant gas accretion take place.  This means that
the distribution of planetary masses is effectively limited by the surface
density of the solid disc. The radial extent of the planet population is at
least a factor of ten smaller than around solar-mass stars and we find that
the peak of the distribution shifts to significantly smaller radii as the
stellar mass decreases, reflecting the fact that disc densities at any given
radii are lower around brown dwarfs.

Whilst the lack of gas accretion limits the predicted mass of the planets, it
also has implications for planetary migration. If a planet does not accrete
gas and remains relatively low mass, it is highly unlikely to open up a gap in
the gaseous disc, and hence will not under-go Type-II migration. This means
that the only migration mechanism relevant for the brown dwarf regime will be
Type-I (which has not been included here). Thus, any observations of planets
around brown dwarfs could shed light directly on the effects of Type-I
migration and hence illuminate its likely character in planet formation around
all higher mass stars.

Note also the paucity of planets plotted in Fig. \ref{FIG:BD1}, indicating
that Earth-like planet formation is highly unlikely in this case. Indeed, we
found a planet with mass $M>0.3M_{\oplus}$ in only $0.035\%$ of the
simulations that we have performed.

\subsection{Impact of Parameter Variation on Planet
  Formation}\label{parameter}

\subsubsection{Surface Density Dependence on Semi-Major Axis}\label{Surface
  Density Dependence on Semi-Major Axis}

It was assumed in section \ref{BD} that the disc profile varied as
$\Sigma\propto a^{-3/2}$. In this section we explore the effect of having a
shallower density profile, with $\Sigma\propto a^{-1}$, while keeping the
scaling with stellar mass as $ \left(M_{\star}/M_{\odot}\right)^2$, and
allowing only Type II migration.  We find that at both solar mass and lower
masses of $M_{\star}=0.05M_{\odot}$, the effect of moving to $k=-1$ is
significant.

\begin{figure}
  \includegraphics[origin=c,width=\columnwidth,trim=0 0 3
  3,angle=0,clip=true]{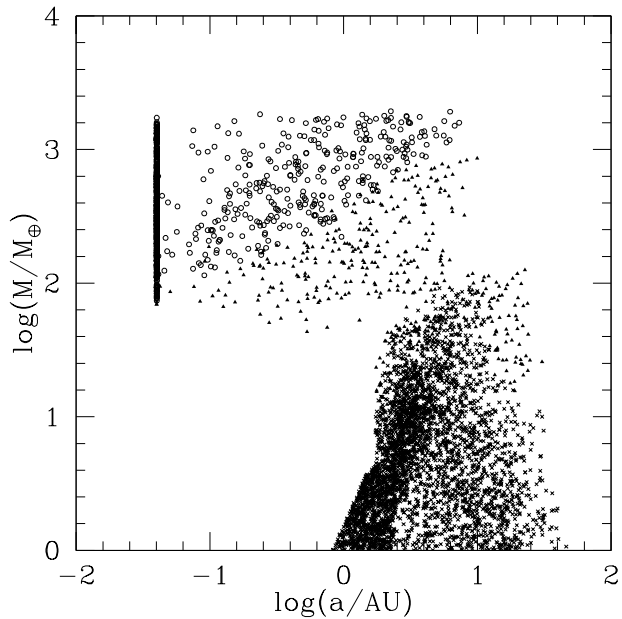}
  \caption{Planet Masses $> 1M_{\oplus}$ for $M_{\star}=M_{\odot}$ with $k = -1$.
    Labels are as in Fig. \ref{FIG:ILcomparisonEarth}.}
  \label{FIG:LargeKEffect1}
\end{figure}

\begin{figure}
  \includegraphics[origin=c,width=\columnwidth,trim=0 0 3
  3,angle=0,clip=true]{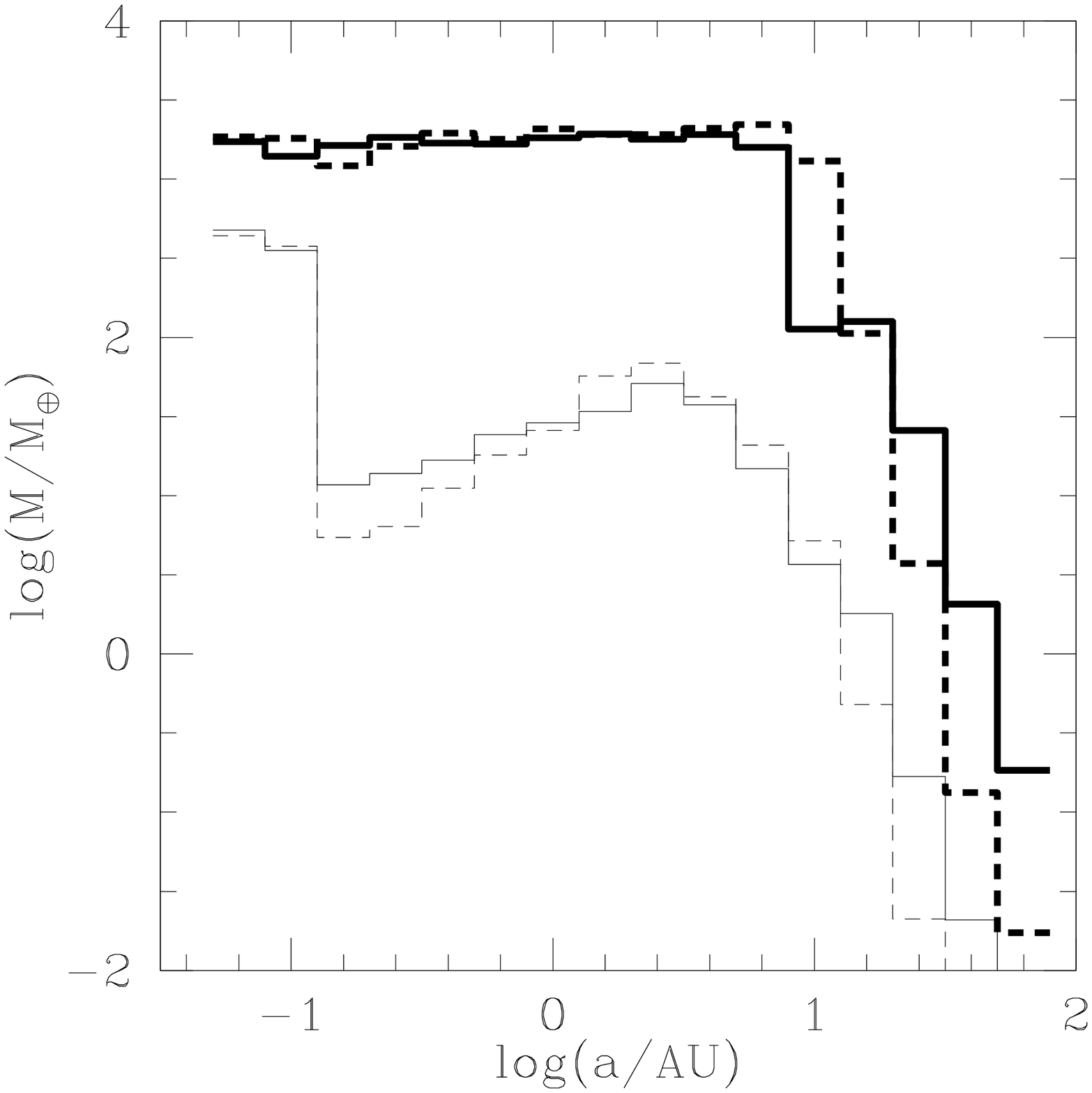}
  \caption{Changing the slope of the surface density profile at
    $M_{\star}=1M_{\odot}$. The thin lines indicate \emph{average} masses, the
    thick \emph{maximum} masses, whilst solid lines indicate $k=-1$, dotted
    lines $k=-3/2$.}
  \label{FIG:LargeKEffect2}
\end{figure}

At $M_{\star}=M_{\odot}$, moving to a $k=-1$ model leaves the average and
maximum \emph{disc} masses unchanged from the standard model in section
\ref{Planet Formation around Solar-Mass Stars}. However, a comparison between
Fig.  \ref{FIG:LargeKEffect1} (which shows the planet mass - semi-major axis
distribution in this case) and Fig. \ref{FIG:ILcomparisonEarth} shows that for
semi-major axes $<10 \textrm{AU}$, the change in model gives very similar
results for the distribution of Jupiter-mass planets, but suppresses the
production of terrestrial planets. Fig. \ref{FIG:LargeKEffect2} shows the
average and maximum planetary masses at different radii for the various cases.
This shows that both the maximum and average planet mass inside
$10\textrm{AU}$ (which are dominated by the gas giants) are similar for these
models, but beyond $10\textrm{AU}$, the effect of the $k=-1$ model is to
increase both the average and maximum plant masses. This result can be simply
understood based on the fact that, for a given disc mass, with a shallower
surface density profile the disc density is decreased at small radii and
increased at large radii.  Note that \citet{2005ApJ...632..670R} looked at
terrestrial planet formation at small radii and found it was suppressed in
discs with shallower profiles, in agreement with our result.

\begin{figure}
  \includegraphics[origin=c,width=\columnwidth,trim=0 0 3
  3,angle=0,clip=true]{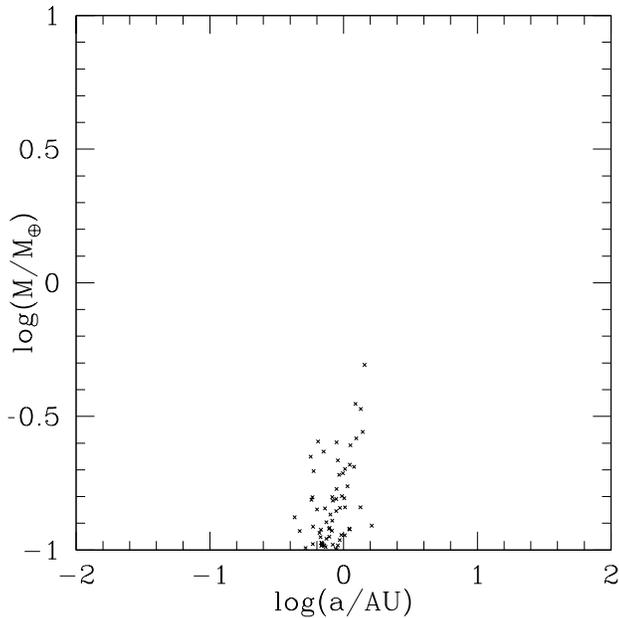}
  \caption{Planet Masses versus semi-major axis for $M_{\star}=0.05 M_{\odot}$
    with $k = -1$. Labels are as in Fig. \ref{FIG:ILcomparisonEarth}.}
  \label{FIG:SmallKEffect1}
\end{figure}

\begin{figure}
  \includegraphics[origin=c,width=\columnwidth,trim=0 0 3
  3,angle=0,clip=true]{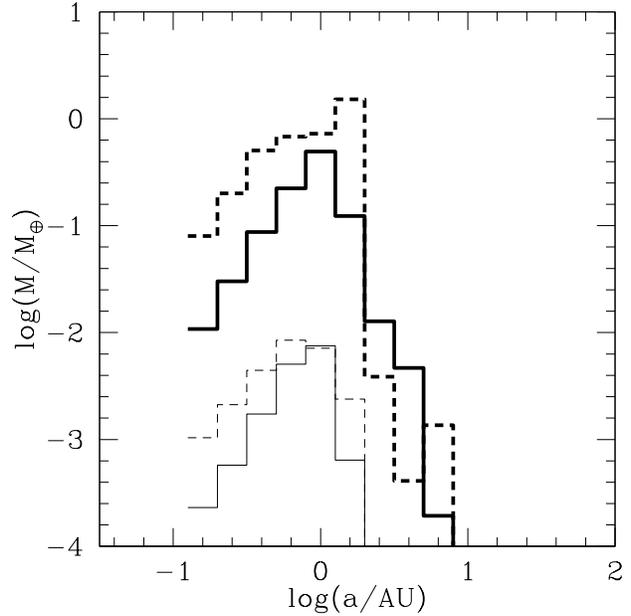}
  \caption{Changing the slope of the surface density profile at
    $M_{\star}=0.05M_{\odot}$. The thin lines indicate {average} masses, the
    thick {maximum} masses, whilst solid lines indicate $k=-1$, dotted lines
    $k=-3/2$.}
  \label{FIG:SmallKEffect2}
\end{figure}

Let us now turn to analyse the case of brown dwarf. Fig.
\ref{FIG:SmallKEffect1} shows the planet mass - semi-major axis distribution
for $k=-1$ (cf. Fig. \ref{FIG:BD1}). The effect of moving to a $k=-1$ model,
similarly to the case of higher stellar mass, is to suppress the formation of
rocky planets. This is further evidenced in Fig. \ref{FIG:SmallKEffect2},
which shows a comparison of the maximum and average planet mass at different
radii for the two cases $k=-1$ (solid line) and $k=-3/2$ (dashed line). Unlike
the case of Solar-type stars, these histograms are now dominated by rocky
planets, and indeed moving to a shallower surface density profiles results in
a reduction of both the maximum and average mass planet formed, preventing the
formation of any planets with masses greater than $1M_{\oplus}$.

\subsubsection{Surface Density Dependence on Stellar Mass}\label{Surface
  Density Dependence on Stellar Mass}

Disc masses estimates for brown dwarfs are very uncertain and only known for a
handful of objects \citep{2006ApJ...645.1498S}. In our standard model,
(section \ref{BD}) we have assumed that protostellar disc masses scale with
stellar mass as $M_{\star}^2$, leading to typical disc masses of a tenth of a
Jupiter mass in the brown dwarf regime. Here, we examine the effect of
allowing for relatively more massive discs around brown dwarfs, and we thus
consider the case where the disc surface density scales linearly with stellar
mass. 

\begin{figure*}
  \centerline{\psfig{figure=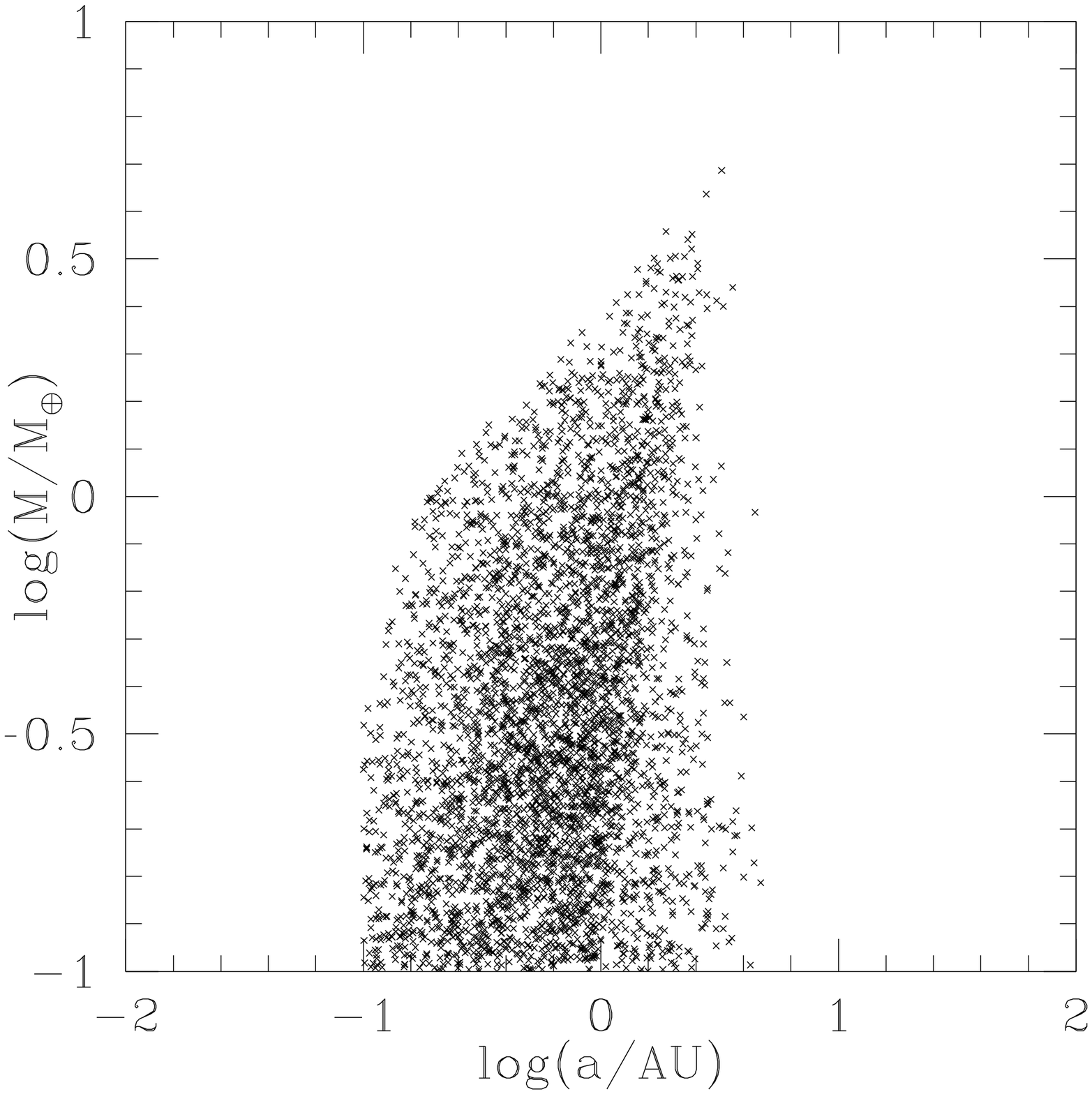,
    width=0.5\textwidth}
    \psfig{figure=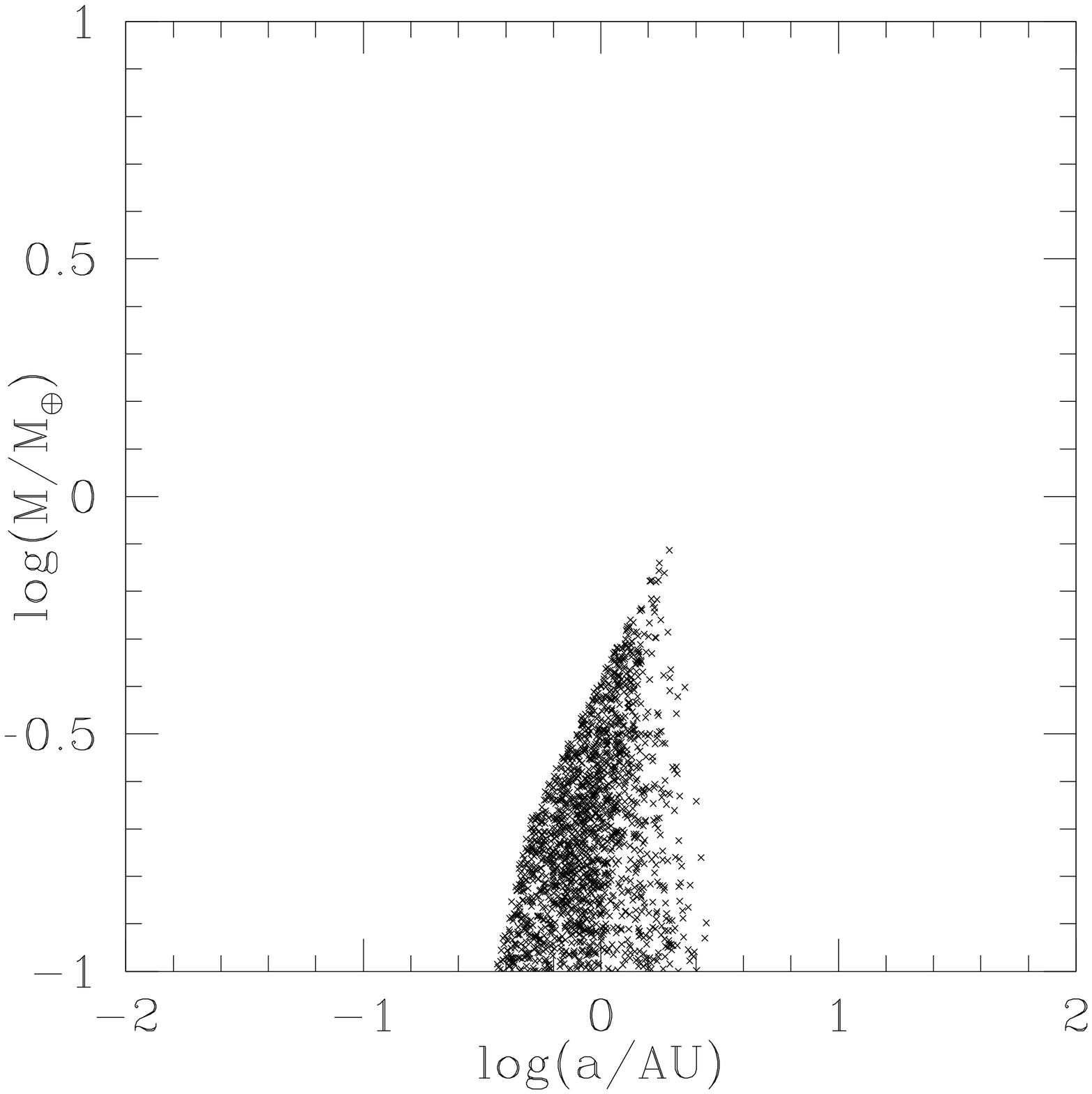,width=0.5\textwidth}}

  \caption{Planet Masses versus semi-major axis for Brown Dwarf with
    $M_{\star}=0.05 M_{\odot}$, $n=1$ and $k=-3/2$ (left panel) and $k=-1$
    (right panel). Labels are as in Fig. \ref{FIG:ILcomparisonEarth}.}
  \label{FIG:BD1:1:A}
\end{figure*}

\begin{figure*}

 \centerline{\psfig{figure=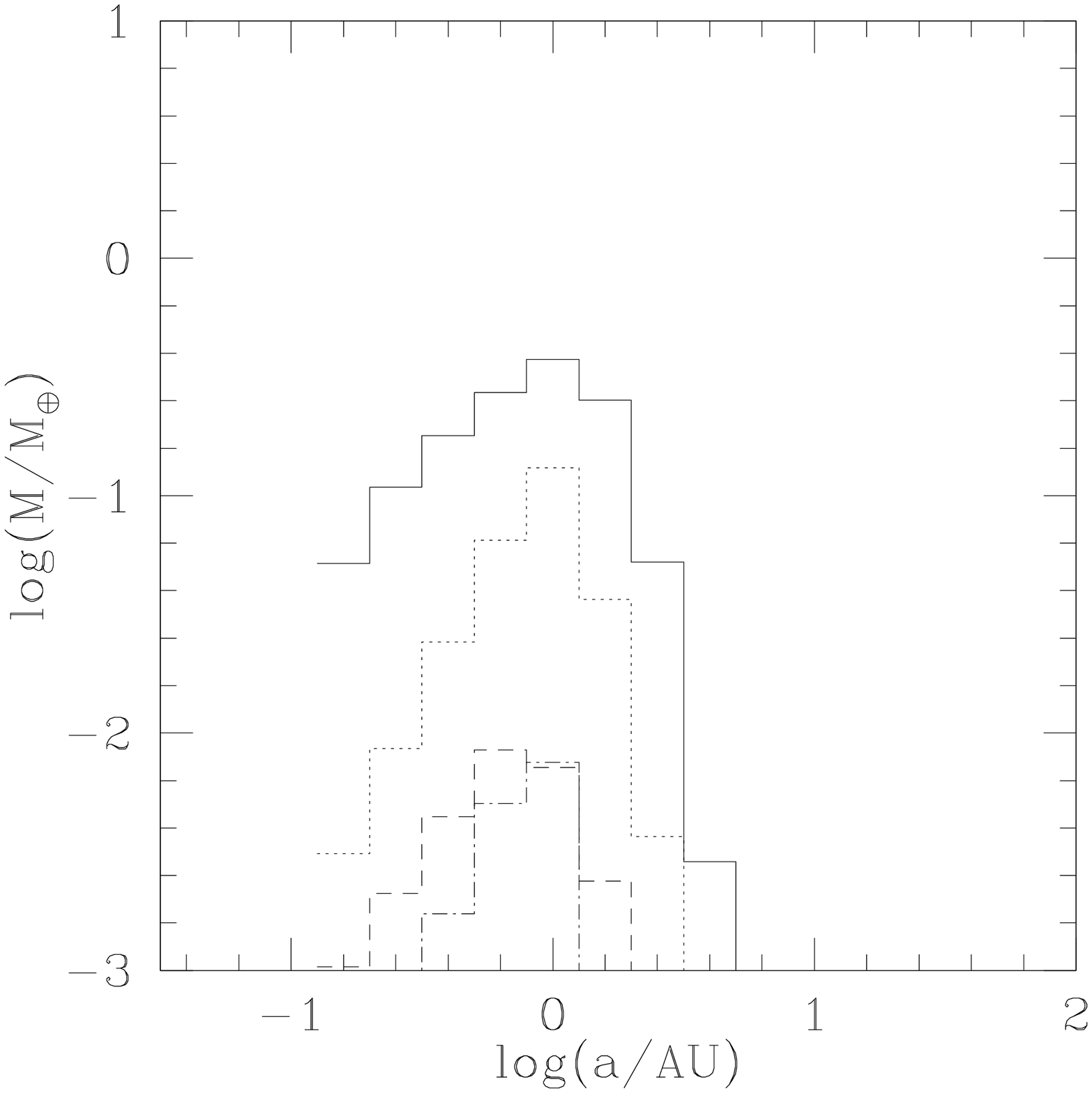,width=0.5\textwidth}
             \psfig{figure=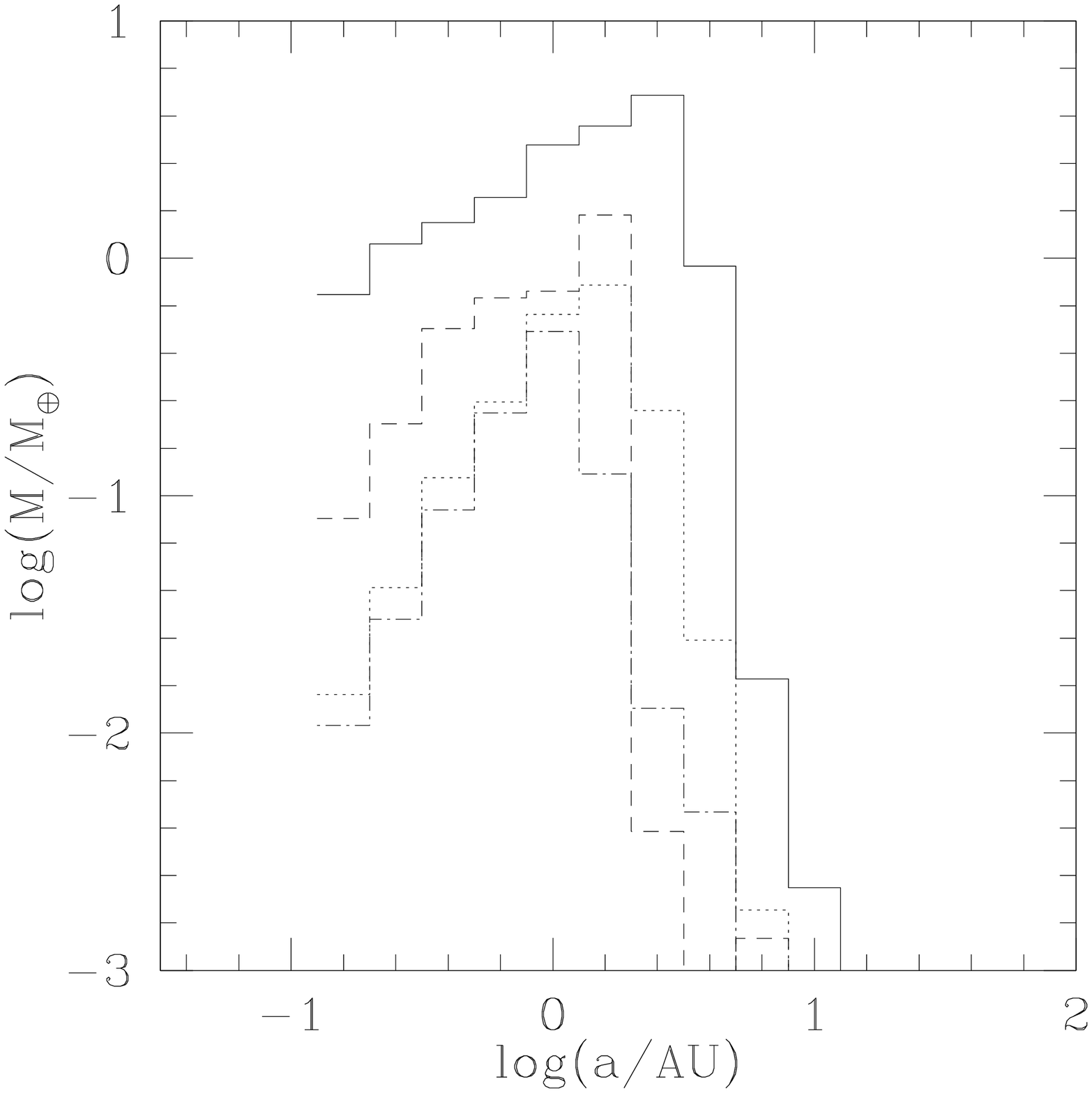,width=0.5\textwidth}
} 
\caption{Results for planet formation around brown dwarfs. The left panel
  shows the average planet masses while the right panel shows the maximum
  planet mass in the various models considered. Solid Line - $n=1$ and
  $k=-3/2$; Dotted Line - $n=1$ and $k=-1$; Dashed Line - $n=2$ and $k=-3/2$;
  Dot/Dash Line - $n=2$ and $k=-1$.}
  \label{FIG:BD1:1}
\end{figure*}

\begin{figure*}

  \centerline{\psfig{figure=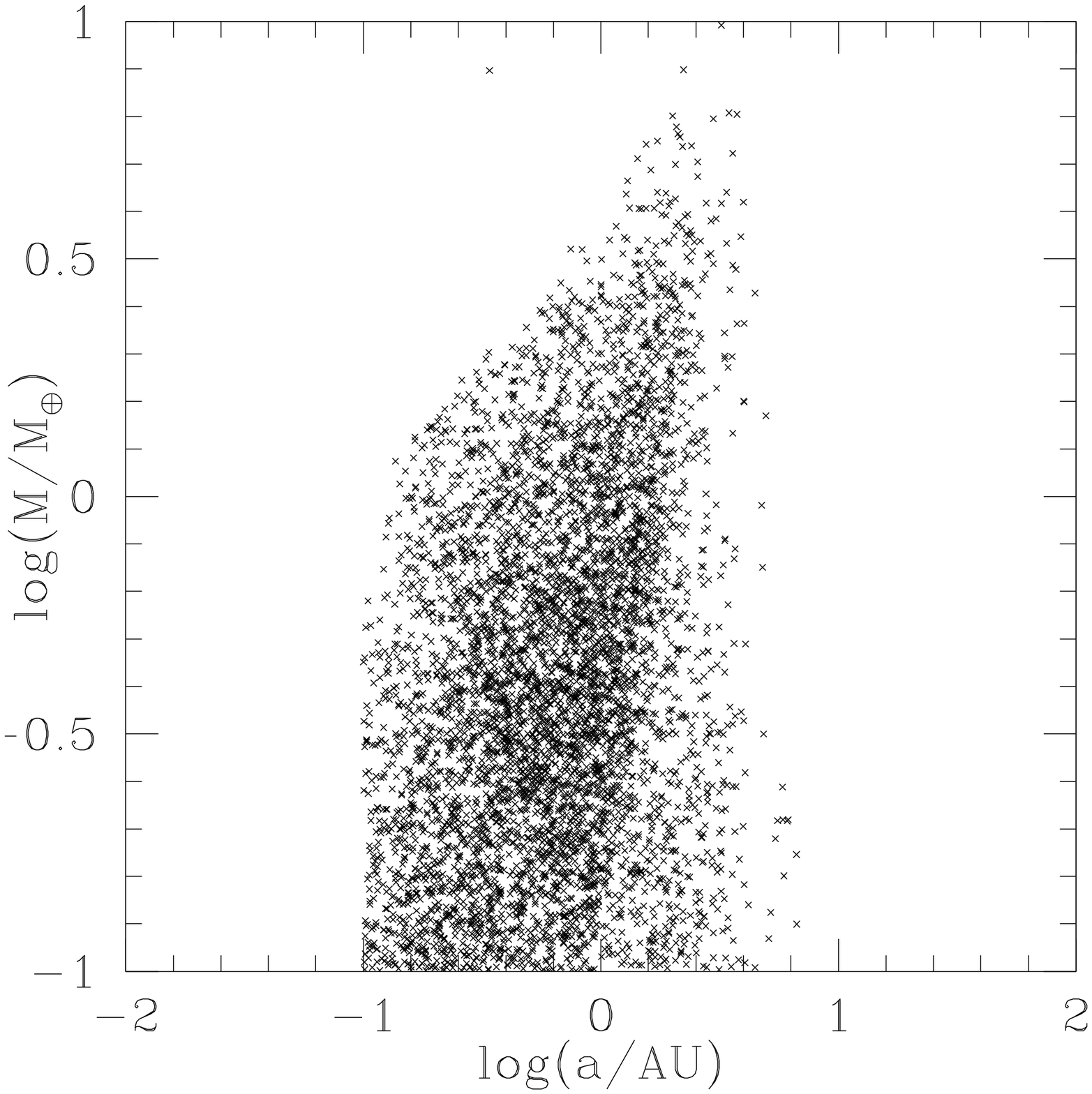,
width=0.5\textwidth}
    \psfig{figure=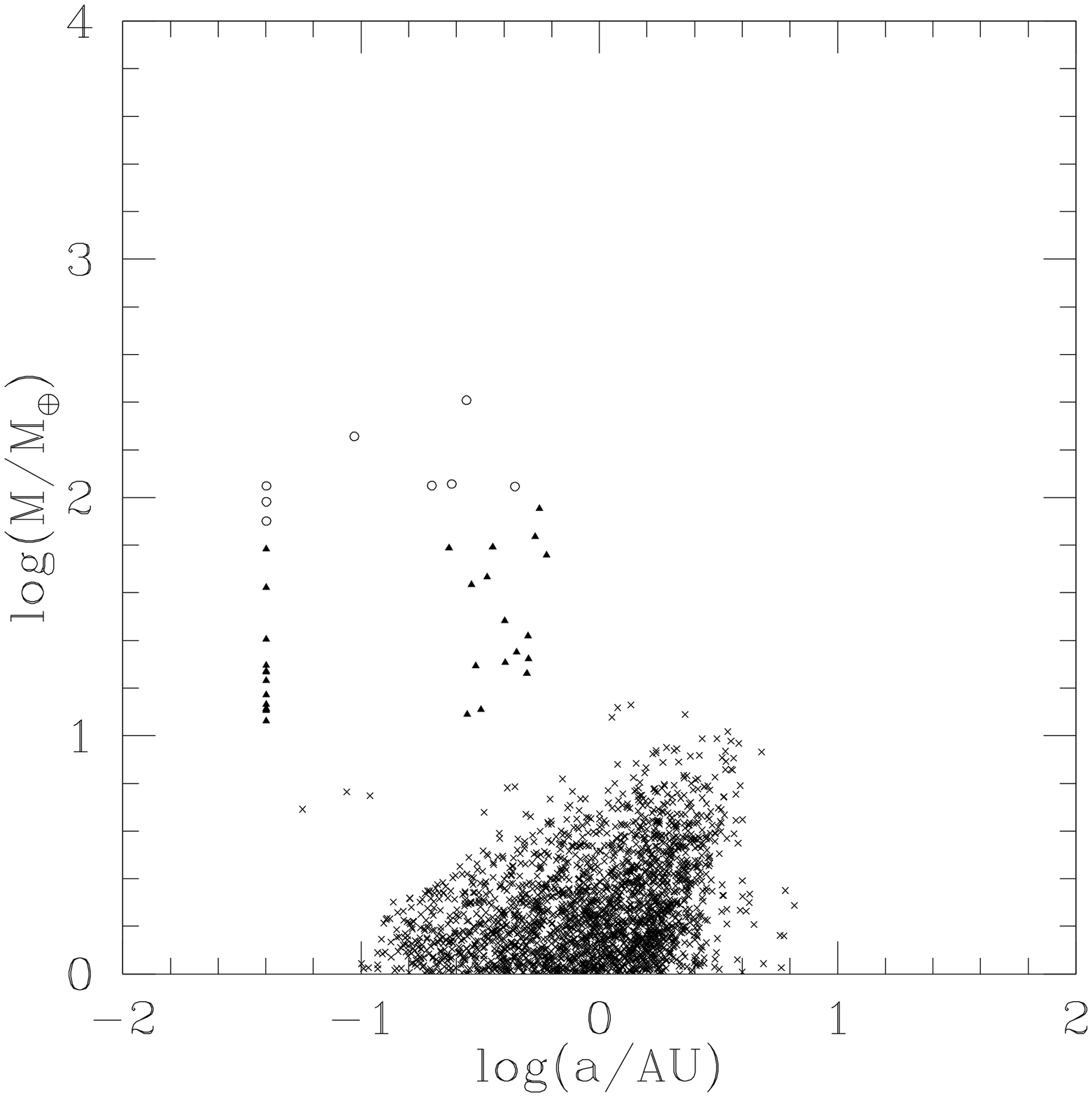,width=0.5\textwidth} }
\caption{Planet mass versus semi-major axis for $M_{\star}=0.05 M_{\odot}$,
  $n=1$ and $k=-3/2$, with outer disc radius equal to 50 AU (left panel) and
  20 AU (right panel). While in the case of $R_{\rm out}=50$ AU, the
  final distribution of planet mass is very similar to the case of $R_{\rm
    out}=100$ AU, some gas giants appear in the case $R_{\rm out}=20$
  AU. These latter cases, however, correspond to the very unlikely case of a
  massive disc ($M_{\rm disc}\gtrsim$ a few Jupiter masses) with a very small
  outer radius.}
  \label{fig:smallsize}
\end{figure*}

We plot in Fig. \ref{FIG:BD1:1:A} the resulting planet mass versus semi-major
axis distribution in this case, for the two cases where $k=-3/2$ (steeper
density profile; left panel) and $k=-1$ (shallower density profile; right
panel) and collate in Fig.  \ref{FIG:BD1:1} the histograms of average and
maximum planet masses formed around brown dwarfs for the various models.

In this case, typical disc masses are of the order of a few Jupiter masses and
the enhanced disc mass is reflected in a larger fraction of objects being able
to form Earth-like planets. Indeed, planets with mass larger than
$0.3M_{\oplus}$ are formed now in $\approx $10\% of the cases for $k=-3/2$ and
in $\approx$ 2\% of the cases for $k=-1$.

In this case, the cores are growing faster, but still none of them grow
quickly enough to start accreting gas before it dissipates.  Thus, despite the
increased potential for planet formation, no object is able to accrete a
significant gaseous envelope, and the maximum planet mass is only slightly
enhanced, the maximum now being $5M_{\oplus}$, in the most favourable case
(that is, $k=-3/2$).

The maximum planetary mass obtained is also slightly dependent on the upper
limit on the disc mass in order for the disc to be gravitationally stable. For
example, if we only allowed discs less massive than 10\% of the central brown
dwarf mass, the maximum resultant planetary mass would be decreased to 1
$M_{\oplus}$ and it would be raised to $9M_{\oplus}$, with some very limited
gas accretion occurring, if the discs are allowed to be half as massive as the
brown dwarf.

Clearly, an important parameter in the determination of the final planet
population is the disc surface density. For a given total disc mass, a more
compact disc, with an outer radius significantly smaller than the $100$ AU
assumed above, would have a larger surface density and therefore an enhanced
chance of forming planets. We have thus re-run the most favourable case for
planet formation (that is, the one with $n=1$ and $k=-3/2$), assuming the same
distribution of total disc mass, but with significantly smaller disc sizes
(this, in turn, implies increasing the factor $f_0$ in Eqs. (1) and (2)). The
results are shown in Fig. \ref{fig:smallsize}, for an outer disc radius
$R_{\rm out}= 50$ AU (left panel) and 20 AU (right panel). While the reduction
of $R_{\rm out}$ to 50 AU only results in a slightly increased maximum planet
mass, in the case of $R_{\rm out}=20$ AU we do see the appearance of a number
of gas giants in a limited number of cases. However, we regard these cases as
extremely unlikely, since they correspond to the upper tail of the disc mass
distribution, with disc masses of the order of several Jupiter masses,
confined within a region as small as 20 AU.

\subsubsection{Introduction of luminosity evolution}
\label{sec:lum}
As mentioned above, the models that we use for planet formation around brown
dwarfs also include the effect of the evolution of the (sub)-stellar
luminosity with time. The stellar luminosity has two main effects on our
models: (a) it determines the temperature of the gas and (b) it sets the
location of the snow-line, inside which the solid density is significantly
reduced (we neglect here any other possible effect related to changes in the
efficiency of angular momentum transport within the discs). Since there is
little or no gas accretion on proto-planets around brown dwarfs in our model,
and no planet becomes massive enough to open up a gap and start migrating, the
first of the above effects actually has essentially no impact on the final
outcome. The second effect is indeed present, but it turns out not to affect
deeply the final results, and simulations performed without taking into
account the luminosity evolution of the brown dwarfs only show relatively
small differences.

\subsubsection{Introduction of planetary migration}\label{migration}

As \citetalias{2005ApJ...626.1045I} have already noted, since low-mass cores
do not undergo Type II migration, any detection of close-in super-Earths
around low mass stars, as in the case of Gl-581-c (a $5.5M_{\oplus}$ planet
orbiting a $0.3M_{\odot}$ star, \citealt{2007arXiv0704.3841U}), would need the
inclusion of other types if migration. In our models of planetary formation
around brown dwarfs in which Type I migration is neglected, we always predict
a peak in the distribution at $1$AU, irrespective of the other parameter
variations in the model.

\begin{figure}
  \includegraphics[origin=c,width=\columnwidth,trim=0 0 0
  0,angle=0,clip=true]{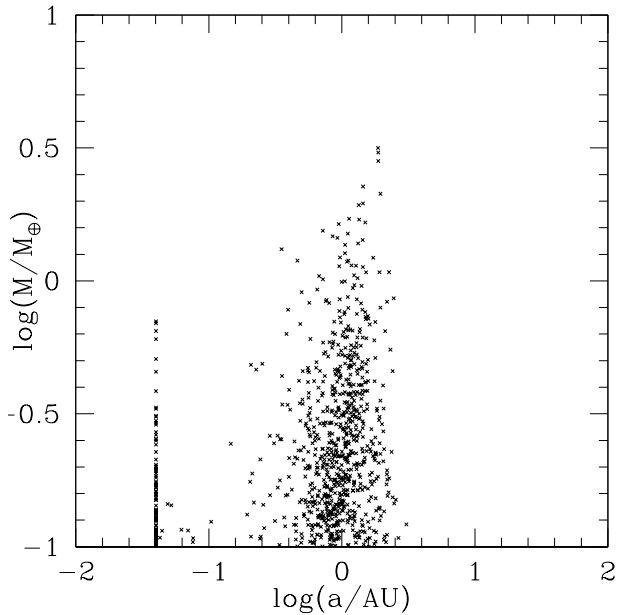}
  \caption{Effect of analytic Type I migration on brown dwarfs with
    $M_{\star}=0.05M_{\odot}$ and $n=1$, $k=-3/2$.  Labels are as in Fig.
    \ref{FIG:ILcomparisonEarth}.}
  \label{FIG:SmallMig}
\end{figure}

As discussed in section \ref{Planet Formation around Intermediate-Mass Stars},
we now consider the addition of a form of Type I migration to our model. In
particular, we consider the model for Type-I migration as derived by
\citet{2002ApJ...565.1257T} for a 3-dimensional isothermal gaseous disc, where
the migration rate is of the form
\begin{eqnarray}
  \tau_I &\approx & 1 \times 10^5 \left(\frac{M}{M_{\oplus}}\right)^{-1} 
  \left(\frac{M_{\star}}{M_{\odot}}\right)^{3/2}
  \left(\frac{\Sigma_{g}}{2 \times 10^3\mbox{g cm}^{-2}}\right)^{-1}
  \textrm{yrs}, 
\label{EQN:T1:TIMESCALE}
\end{eqnarray}
We find that in general the effects of Type I migration around brown dwarfs
are less significant than around solar type stars. This is mostly due to the
fact that planetary cores around brown dwarfs grow at a much slower rate, and
therefore the migration timescale in this case is relatively longer.  For the
$n=1$ $k=-3/2$ model (see Fig. \ref{FIG:SmallMig}), the maximum planet mass at
$\sim 1\textrm{AU}$ is very slightly reduced, whilst the average value at this
radius (and the efficiency of the planet formation process) is reduced as a
large number of the cores are swept into the inner boundary.

Since Type-II migration effectively never occurs around brown dwarfs, any
observational evidence of a secondary peak/excess in the radial distribution
of planetary semi-major axes inside the main distribution peak at
$\sim1\textrm{AU}$ would provide a clean, direct insight into the nature and
extent of Type-I migration.

\section{Discussion and Conclusions}\label{Discussion and Conclusions}

We have extended standard core-accretion models for planet formation down to
the brown dwarf regime and investigated the potential for planet formation in
protoplanetary discs around objects with masses $\sim 0.05M_{\odot}$. We
looked at the impact on the predicted planetary mass - semi-major distribution
of varying fundamental parameters, like the total disc mass and its radial
distribution.

In line with previous results \citepalias{2004ApJ...604..388I}, at higher
stellar masses ($M_{*} \sim 1M_{\odot}$, Fig. \ref{FIG:ILcomparisonEarth}), we
found that the full range of extant planet types is produced, including rocky
and icy cores up to $10^2M_{\oplus}$, rock and ice giants spanning a range
$10^1 - 10^3 M_{\oplus}$ and gas giants extending from $10^2 M_{\oplus}$ up to
$10^4M_{\oplus}$. In addition, we have found that a shallower surface density
profile results in a reduced efficiency of the formation of rocky planets at
radii $\lesssim 10$ AU.

Our main result is the determination of the likelihood of planet formation
around brown dwarfs, with mass $\sim 0.05M_{\odot}$.  Our main findings can be
summarized as follows. ({\it i}) Giant planet formation is completely
inhibited in this case. In none of our standard simulations did any planet
accrete a significant gaseous envelope. This occurs despite the fact that in a
few cases relatively massive cores do form (in principle massive enough to
start runaway gas accretion), because such massive cores, when they form, only
do so at a late time, when most of the gaseous component of the disc has
dissipated. This means that Jupiter mass companions to brown dwarfs, as
sometimes observed, have to be formed through a different -- binary like --
mechanism, that is either core or disc fragmentation. While this had already
been discussed for very wide extremely low mass binaries, like 2MASS1207
\citep{2005MNRAS.364L..91L}, we can now extend this conclusion to closer
binaries. ({\it ii}) The formation of Earth-like planets is possible even
around brown dwarfs, and the maximum planet mass that we have found is
$\approx 5M_{\oplus}$. However the likelihood of this process is critically
dependent on what the typical disc masses are in this regime. In particular,
if typical brown dwarf disc masses are of the order of a few Jupiter masses
(which is consistent with a linear scaling of disc mass with stellar mass),
then up to 10\% of brown dwarfs might possess planets with masses $>0.3
M_{\oplus}$, while on the other hand if typical brown dwarf disc masses are
only a fraction of a Jupiter mass (consistent with a quadratic scaling of disc
mass with stellar mass), then even the formation of Earth-like planets would
be essentially prohibited. We note that current estimates of brown dwarfs disc
masses are only available for a small number of objects and are very close to
our boundary between planet-forming and non-planet-forming systems, with
upper limits at the level of a few Jupiter mass \citep{2006ApJ...645.1498S}.
In order to answer the question of the occurrence of planets around brown
dwarfs it is therefore essential to improve the statistics and the accuracy of
disc mass measurements in the very low mass regime.

Finally, we note that the lack of any gas accretion in the planet formation
process around brown dwarfs and the concomitant lack of any Type-II migration
means that the radial distribution of the planets will be determined purely by
the initial mass distribution in the disc and the subsequent effects of Type-I
migration. Given that the underlying disc models tend to produce the most
massive planets at $a\sim 1 \textrm{AU}$, irrespective of the specific model,
then any excess/secondary peak (or lack thereof) in the number of detected
planets inside this radius would give clear evidence of the nature of Type-I
migration.

\section*{Acknowledgments}\label{Acknowledgements}
We would like to thank Daniel Apai, Cathie Clarke, Ilaria Pascucci, Doug Lin,
Jim Pringle and Chris Tout for helpful discussions.  MJP acknowledges the UK
PPARC/STFC for a research studentship.

\bibliography{references}
\bibliographystyle{mn2e}

\end{document}